\documentclass[11pt,a4paper]{article}
\usepackage[pdftex]{color}
\usepackage{a4wide,amssymb,amsmath,latexsym,epsfig,epstopdf}
\usepackage{verbatim}
\usepackage[numbers]{natbib}

\newtheorem{lemma}{Lemma}

\newtheorem{theorem}{Theorem}
\newtheorem{proposition}{Proposition}

\newenvironment{proof}{{\it Proof:\/}}{\hspace*{1em}\hfill$\Box$\vskip 0.1in}

\newenvironment{myproof}{\begin{proof}}{\end{proof}}
\newcommand{\SRPT}{\textsc{Shortest Remaining Processing Time First}}
\newcommand{\SMITH}{\textsc{Smith Ratio Algorithm}}
\newcommand{\CONSV}{\textsc{Conservative Algorithm}}
\newcommand{\EDF}{\textsc{Earliest Deadline First}}

\newcommand{\ALG}{\textsc{Alg}}
\newcommand{\EXPRI}{\textsc{Exponential Capacity Algorithm}}

\title{Online Scheduling of Bounded Length Jobs to Maximize Throughput}

\author{
Christoph \textsc{D\"urr}%
\thanks{CNRS, LIX UMR 7161, Ecole Polytechnique
  91128 Palaiseau, France.
  Supported by ANR Alpage.}
\and \L{}ukasz \textsc{Je\.z}%
\thanks{Institute of Computer Science,
  University of Wroc{\l}aw,
  50-383 Wroc{\l}aw, Poland.
  Supported by MNiSW grant N N206 1723 33, 2007--2010,
  and COST 295 ``DYNAMO''}
\and \textsc{Nguyen Kim} Thang\footnotemark[1]%
}

\begin{document}

\maketitle

\begin{abstract}
  We consider an online scheduling problem, motivated by the issues
  present at the joints of networks using ATM and TCP/IP. Namely, 
  IP packets have to broken down to small ATM cells and sent out
  before their deadlines, but cells corresponding to different packets
  can be interwoven.
  More formally, we consider the online scheduling problem with
  preemptions, where each job $j$ is revealed at release time $r_j$,
  has processing time $p_j$, deadline $d_j$ and weight $w_j$.  
  A preempted job can be resumed at any time.
  The goal is to maximize the total weight of all jobs completed on time.
  Our main result are as follows: we prove that if all jobs have processing
  time \emph{exactly} $k$, the deterministic competitive ratio is between
  $2.598$ and $5$, and when the processing times are \emph{at most} $k$,
  the deterministic competitive ratio is $\Theta(k/\log k)$.
\end{abstract}

\section{Introduction}

Many Internet service providers use an ATM network which has been
designed to send telephone communication and television broadcasts, as
well as usual network data.  However, the Internet happens to use
TCP/IP, so at the joints of these networks IP packets have to be
broken down into small ATM cells and fed into the ATM network. This
raises many interesting questions, as ATM network works with fixed
sized cells (48 bytes), while IP network works with variable sized
packets.  In general, packet sizes are bounded by the capacity of
Ethernet, i.e. 1500 bytes, and in many cases they actually achieve
this maximal length. Ideally packets also have deadlines and
priorities (weights).  The goal is to maximise the \emph{quality of
  service}, i.e. the total weight of packets that have been entirely
sent out on time.

This problem can be formulated as an online-scheduling problem on a
single machine, where jobs arrive online at their release times, have
some processing times, deadlines and weights, and the objective is to
maximise the total weight of jobs completed on time.  Preemption is
allowed, so a job $i$ can be scheduled in several separated time intervals,
as long as their lengths add up to $p_i$.  Time is divided into integer
time steps, corresponding to the transmission time of an ATM cell, and all
release times, deadlines and processing times are assumed to be integer.
This problem can be denoted as
$1|\text{online-}r_{i};\text{pmtn}|\sum w_i(1-U_i)$,
according to the notation of~\cite{Chen.Potts.Woeginger:review}.


\subsection{Our results}
In this paper we consider the case when processing times of all jobs
are bounded by some constant $k$, and the case when they equal
$k$. Both variants are motivated by the network application in
mind. We study the competitive ratio as a function of $k$. Our main
results are as follows.
\begin{itemize}
  \item We provide an optimal online algorithm for the bounded
    processing time case, that reaches the ratio $O(k/\log k)$.
  \item 
    We provide a simple $5$-competitive algorithm for the equal
    processing time case.
\item 
    For the same case a $2.59$-lower bound on the competitive ratio
    was stated in
    \cite{ChanLamTing:New-Results-on-On-Demand-Broadcasting} that
    applies also to our model with preemption.  However, the proof is
    incomplete (see discussion at the end of section~\ref{sec:2.598}).
    For completeness we provide a $3\sqrt{3}/2 \approx 2.598$ lower
    bound on the competitive ratio.
\end{itemize}

In addition we also provide several minor results, some of which are moved
to the appendix due to space constraints.
\begin{itemize}
\item
  For the bounded processing time case, we show that the well-known
  \SMITH{} is $2k$-competitive, and provide an example tight up to
  a factor of $2$.
  We also show that asymptotically the competitive ratio of any
  deterministic algorithm is at least $k/\ln k$, improving the previous
  bound~\cite{Ting:broadcasts} of $k/(2\ln k) -1$ by a factor of $2$.
\item
  For bounded processing time with unit weights, it is known that the
  competitive ratio is $\Omega(\log k/\log\log k)$ when time points
  are allowed to be
  rationals~\cite{BaruahHaritsaSharma:On-line-scheduling-to-maximize}.
  We provide an alternative proof for the more restricted integer variant,
  obtaining better multiplicative constant at the same time.
\item
  It was previously
  stated~\cite{KalyanasundaramPruhs:Maximizing-job-completions-online}
  that \SRPT{} is $O(\log k)$-competitive for the bounded processing
  time, unit weight model. This result follows from a larger proof.
  For completeness, we provide a concise proof that \SRPT{} is
  $2H_k$-competitive.
\end{itemize}

\subsection{Related work}

It is known that the general problem without a bound on processing
times has an unbounded deterministic competitive
ratio~\cite{BaruahHaritsaSharma:On-line-scheduling-to-maximize}, so
different directions of research were considered.  One is to see if
randomisation helps, and indeed in~%
\cite{KalyanasundaramPruhs:Maximizing-job-completions-online} a
constant competitive randomized algorithm was given, although with a
big constant.  Another direction of research is to consider resource
augmentation, and in~\cite{Koo.Lam.ea:tight-deadlines} a deterministic
online algorithm was presented, which has constant competitive ratio
provided that the algorithm is allowed a constant speedup of its
machine compared to the adversary.  Finally a third direction is to
restrict to instances with bounded processing time.

\begin{description}
  \item[Bounded processing time, unit weights] (Case $\forall
    j:p_j\leq k,w_j=1$) The offline problem can be solved in time
    $O(n^4)$~\cite{Baptiste:n4} already when the processing time is
    unbounded. Baruah et
    al.\ \cite{BaruahHaritsaSharma:On-line-scheduling-to-maximize}
    showed that any deterministic online algorithm is $\Omega(\log
    k/\log\log k)$-competitive in a model where processing times,
    release times and deadlines of jobs can be rational.  The
    currently best known algorithm is \SRPT{}, which is $O(\log
    k)$-competitive
    \cite{KalyanasundaramPruhs:Maximizing-job-completions-online}.
    The same paper provides a constant competitive randomized
    algorithm, however with a large constant.
  \item[Bounded processing time, arbitrary weights] (Case $\forall
    j:p_j\leq k$) For fixed $k$ the offline problem has not been
    studied to our knowledge, and when the processing times are
    unbounded the offline problem is $\mathcal{NP}$-hard by a trivial
    reduction from Knapsack Problem.  It is known that any
    deterministic online algorithm for this case has competitive ratio
    $k/(2\ln k) -1$ \cite{Ting:broadcasts}. For the variant with only
    tight jobs, Canetti and Irani \cite{Canetti.Irani:randomized}
    provide an $O(\log k)$-competitive randomized online algorithm
    and show a $\Omega(\sqrt{\log k/\log\log k})$ lower bound for any
    randomized competitive algorithm against an oblivious adversary.
 \item[Equal processing time, unit weights] (Case $\forall
    j:p_j=k,w_j=1$) The offline problem can be solved in time
    $O(n\log n)$~\cite{Lawler:Knapsack-like-Scheduling-Problems}, and
    it is well known that the same algorithm can be turned into a
    1-competitive online algorithm, see for
    example~\cite{Vakhania:lawler}.
 \item[Equal processing time, arbitrary weights] (Case $\forall j:
    p_j=k$) The offline problem can be solved in time
    $O(n^4)$~\cite{Baptiste.Chrobak.ea:throughput}. 
    For $k=1$ the problem is well studied, and the
    deterministic competitive ratio is between 1.618 and 1.83 
    \cite{Li.Sethuraman.Stein:buffer, EngWes07}.
\end{description}
Our model is sometimes called the \emph{preemptive model with resume},
as opposed to
\emph{preemptive model with restarts}~\cite{Chrobak.Jawor.ea:restarts},
in which an interrupted job can only be processed from the very beginning.
\emph{Overloaded real-time systems}%
~\cite{BaruahHaritsaSharma:On-line-scheduling-to-maximize} form another
related model, in which all the job parameters are reals, the time is
continuous, and uniform weights are assumed.

\section{Preliminaries} \label{sec:intro}

For a job $i$ we denote its release time $r_i$, its deadline $d_i$,
its processing time $p_i$ and its weight $w_i$. All these quantities,
except $w_i$, are integers. Let $q_i(t)$ be the remaining processing
time of job $i$ for the algorithm at time $t$. When there is no
confusion, we simply write $q_{i}$. We say that job $i$ is
\emph{pending} for the algorithm at time $t$ if it has not been
completed yet, $r_i\leq t$, and $t+q_i(t)<d_i$.  Let $j$ be a job
uncompleted by the algorithm.  The \emph{critical time} of $j$ is the
latest time when $j$ was still pending for the algorithm. In other
words, the critical time $s$ of job $j$ for the algorithm is such
moment $s$ that if the algorithm does not schedule $j$ at time $s$, it
cannot finish $j$ anymore, i.e.  $s =\max\{\tau: \tau + q_{j}(\tau) =
d_{j}\}$.
We assume that a unit $(i,a)$ scheduled at time $t$ is processed during
the time interval $[t,t+1)$, i.e. its processing is finished just before
time $t+1$. For this reason by \emph{completion time} of a job $i$ we mean
$t+1$ rather than $t$, where $t$ is the time its last unit was scheduled.

Throughout the paper we analyse many algorithms with similar charging
schemes sharing the following outline: for every job $j$ completed by
the adversary we consider its $p_j$ units.  Each unit of job $j$ will
charge $w_j/p_j$ to some job $i_0$ completed by the algorithm.  The
charging schemes will satisfy the condition that every job $i_0$
completed by the algorithm receives a total charge of at most $R
w_{i_0}$, which implies $R$-competitiveness of the algorithm.

More precisely we distinguish individual units scheduled by both the
algorithm and the adversary, where unit $(i,a)$ stands for execution
of job $i$ when its remaining processing time was $a$.  In particular
a complete job $i$ consists of the units $(i,p_i), (i,p_i-1),
\ldots,(i,1)$.  With every algorithm's unit $(i,a)$ we associate a
\emph{capacity} $\pi(i,a)$ that depends on $w_i$ and $a$, whose exact
value will be different from proof to proof. The algorithms, with
their capacities, will be designed in such a way that they satisfy the
following properties, with respect to $\pi$.
\begin{description}
  \item[$\rho$-monotonicity:] If the algorithm schedules
    $(i,a)$ at $t$ and $(i',a')$ at $t+1$ with $a>1$, then
    $\rho \pi(i',a')\geq \pi(i,a)$,
  \item[validity:] If a job $j$ is pending for the algorithm at any time $t$,
    then the algorithm schedules a unit $(i,a)$ at $t$ such that 
    $\pi(i,a) \geq w_{j}/p_{j}$.
\end{description}
Let us remark that our algorithms are $\rho$-monotone for some $\rho < 1$.
Also note that if at time $t$ there is a job $j$ pending for a valid algorithm,
the algorithm schedules a unit of some job at $t$.

\begin{figure}[ht]
  \centerline{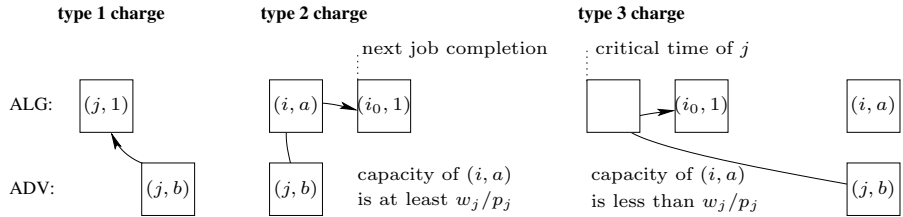}
  \caption{The general charging scheme}
  \label{fig:scheme}
\end{figure}

In general there will be 3 types of charges in the charging scheme; these
are depicted in Figure~\ref{fig:scheme}.
Let $(j,b)$ be a unit of job $j$ scheduled by the adversary at time $t$.
\begin{description}
\item[Type 1:] If the algorithm already completed $j$ by time $t$,
  then charge $w_j/p_j$ to~$j$.
\item[Type 2:] Otherwise if the algorithm schedules a job unit $(i,a)$
  at time $t$ that has capacity at least $w_j/p_j$ then we charge
  $w_j/p_j$ to $i_0$, where $i_0$ is the next job completed by the
  algorithm from time $t+1$ on.
\item[Type 3:] In the remaining case, the job $j$ is not pending
  anymore for the algorithm, by the algorithm's validity.
  Let $s$ be the critical time of $j$.  We charge
  $w_j/p_j$ to $i_0$, where $i_0$ is the first job completed by the
  algorithm from time $s+1$ on. Note that $\pi(i_0,1) \geq
  w_j/p_j$, by validity and monotonicity.
\end{description}

Clearly every job $i_0$ completed by the algorithm can get at most
$w_{i_0}$ type 1 charges in total.  We can bound the other types as
well.
\begin{lemma}                                          \label{lem:type3}
  Let $\cal J$ be the set of job units that are type 3 charged to a
  job $i_0$ completed by a monotone and valid algorithm.
  Then for all $p$ there are strictly less than $p$ units $(j,b) \in \cal J$
  with $p_j \leq p$.  In particular, $|{\cal J}| \leq k-1$ if all jobs have
  processing time at most $k$. Moreover, for each $(j,b) \in \cal J$ it holds
  that $w_j/p_j \leq \pi(i_0,1)$.
\end{lemma}
\begin{myproof}
  To be more precise we denote the elements of $\cal J$ by triplets
  $(s,t,j)$ such that a job unit $(j,b)$ scheduled at time $t$ by the
  adversary is type 3 charged to $i_0$ and its critical time is $s$.
  Let $t_0 \geq s$ be the completion time of $i_0$ by the algorithm. 
  Between $s$ and $t_0$ there is no idle time, nor any other job completion,
  so by monotonicity and validity of the algorithm the
  capacities of all units in $[s,t_0)$ are at least $w_j/p_j$.
    However by definition of type 3 charges, the algorithm schedules
    some unit with capacity strictly smaller than $w_j/p_j$ at $t$, so
    $t_0 \leq t$.
    
    Since $s$ is the critical time of $j$, $s+q_j(s)=d_j$.
    However, since the adversary schedules $j$ at time $t$ we have
    $t<d_j$.  Thus $t-s<q_j(s) \leq p_j$.  Note that all triplets
    $(s,t,j)\in\cal J$ have distinct times $t$.  The first part of the
    lemma follows from the observation that there can be at most $c-1$
    pairs $(s,t)$ with distinct $t$ that satisfy $s\leq t_0 \leq t$ and $t-s<c$.

    Since $j$ was pending at time $s$, the unit scheduled by the algorithm
    at time $s$ had capacity at least $w_j/p_j$.  By monotonicity
    of the algorithm the same holds at time $t_0-1$, so
    $\pi(i_0,1) \geq w_j/p_j$.
\end{myproof}

\begin{lemma}                                          \label{lem:type2}
Let $\rho<1$. Then the total type 2 charge a job $i_0$ completed by a
$\rho$-monotone and valid algorithm receives is at most $\pi(i_0,1) / (1-\rho)$.
\end{lemma}
\begin{myproof}
  Let $t_0$ be the completion time of $i_0$, and let $s$ the smallest time such
  that $[s,t_0)$ contains no idle time and no other job completion.  Then the
  unit scheduled at time $t_0-i$ for $1\leq i \leq t_0-s+1$ has capacity at most
  $\pi(i_0,1)\rho^{i-1}$, by $\rho$-monotonicity.  Thus the total type 2 charge
  is bounded by
\[
    \pi(i_0,1) ( 1 + \rho + \rho^2 + \rho^3 \ldots ) 
  = \pi(i_0,1)  / (1-\rho) \enspace.
\]
\end{myproof}

In the next sections, we adapt this general charging scheme to individual
algorithms, demonstrating that the class of algorithms that can be analysed this
way is very rich. Note that as this is only an analysis framework, one still
needs to design their algorithm carefully, and then appropriately choose the
capacity function. In particular, it is possible to analyse a fixed algorithm
using different capacity functions, and their choice greatly affects the upper
bound on the algorithm's competitive ratio one obtains.

All our algorithms at every step schedule the job with maximum capacity, but
this is not a requirement for the scheme to work.  For example, some of our
preliminary algorithms did not work this way.  We also believe our scheme could 
be adapted to the model with real number parameters, as in the case of
\emph{overloaded real-time systems} for example, even if arbitrary weights
are allowed.



\section{Bounded Processing Times}

This time we consider instances with arbitrary weights.  A natural
algorithm for this model, the \textsc{Smith Ratio Algorithm}, schedules
the pending job $j$ that maximizes the Smith ratio $w_j/p_j$ at every step.
A very simple instance with only two jobs ($r_a=r_b=0$, $p_a=d_a=w_a=k$,
$p_b=1$, $w_b=1+\epsilon$, $d_b=k+1$) shows that its competitive ratio is
no better than $k+1$. It turns out that $2k$-competitiveness can be proved
just as easily using our charging scheme. We give the proof for
completeness, and then introduce an optimal algorithm.
 \begin{theorem}
   The {\SMITH} is $2k$-competitive.
 \end{theorem}
 \begin{myproof}
   We use the general charging scheme.
   The algorithm is $\frac{k-1}{k}$-monotone and valid w.r.t.
   $\pi(i,a)=w_i/a$.  Each job $i_0$ completed by the algorithm receives at
   most $w_{i_0}$ type 1 charge in total.
   Lemma~\ref{lem:type2} implies that each $i_0$ receives at most $k w_{i_0}$
   type 2 charges in total, as for $\pi(i,a)=w_i/a$ the value of $\rho$ is
   $1-1/k$.
   By Lemma~\ref{lem:type3}, $i_0$ receives at most $k-1$ type 3
   charges, and each such charge is at most $\pi(i_0,1)=w_{i_0}$.
   This concludes the proof.
 \end{myproof}

\textsc{The Exponential Capacity Algorithm}
in every step schedules the job $j$ that maximises the value of function
$\pi(j,q_j) = w_j \cdot \alpha^{q_j-1}$. This $\pi$ is in fact the capacity
function we use in the analysis, and $\alpha<1$ is a parameter that we specify
later.

In fact, the constant $\alpha$ depends on $k$, seemingly making
{\EXPRI} semi-online.  However, the $\alpha(k)$ we use is
an increasing function of $k$, and the algorithm can be made fully
online by using the value $\alpha(k^*)$ in each step, where $k^*$ is
the maximum processing time among all jobs released up to that step.
Let $\pi^*$ denote the capacity function defined by $\alpha(k^*)$.
The fully online algorithm is trivially $\alpha(k^*)$-monotone and
valid with respect to $\pi^*$, as both $\alpha(k^*)$ and $\pi^*$ only
increase as time goes.  This allows us to analyse the algorithm using
the final values of $k^*$ and $\pi^*$.
\begin{theorem}
  The {\EXPRI}  is $\left(3+o(1)\right)k/\ln k$-competitive.
\end{theorem}
\begin{myproof}
   As before, we use the general charging scheme. Let us define the
   proper value of $\alpha(k)$ now: $\alpha(k) = 1 -c^2 \cdot \ln
   k/k$, where $c = 1-\epsilon$ for arbitrarily small $\epsilon > 0$.
   The algorithm is clearly $\alpha$-monotone.

   To prove validity it is sufficient to prove that $p\alpha^{p-1} \geq 1$
   for all $p \leq k$, as this implies $w_j/p_j\leq w_j\alpha^{p_j-1}$, and,
   by monotonicity and the choice of $\pi$, the following holds at any time
   step $t$ and job $j$ pending at $t$.
   \begin{equation}
   w_j\alpha^{p_j-1}\leq\pi\left(j,q_j(t)\right)\leq\pi\left(h,q_h(t)\right)
   \leq \pi\left(i_0,1\right) = w_{i_0} \enspace, \label{expri:val-type-3}
   \end{equation}
   where $h$ is the job scheduled by the algorithm at $t$, and $i_0$ is the
   next job completed by it from time $t+1$ on. Hence we introduce the function
   $f(x) = x\alpha^{x-1}$, and claim the following holds for any large enough
   $k$ and any $x \in \{1,2,\ldots,k\}$.
\begin{align}
\label{ExPri: few large}
    f(x) & \geq 1 &\mbox{ for } 1\leq x \leq \frac{k}{c^2 \ln k} \enspace,
\\
\label{ExPri: many small}
    f(x) & \geq \ln k
         &\mbox{ for } \frac{k}{c^2\ln k} < x \leq k \enspace.
\end{align}
   In particular $f(x) \geq 1$ for $x \in \{1,2,\ldots,k\}$, hence the algorithm
   is valid by~(\ref{expri:val-type-3}).

   Now we bound the total charge of type 3 any job $i_0$ can receive.
   Let $\cal J$ denote the set of job units that are type 3
   charged to $i_0$.  For each $(j,b) \in \cal J$ the charge
   from it is $w_j/p_j$, while $w_j\alpha^{p_j-1}\leq w_{i_0}$,
   by~(\ref{expri:val-type-3}).  Thus $w_j/p_j\leq w_{i_0}/(p_j\alpha^{p_j-1})
   = w_{i_0}/f(p_j)$.  Recall that Lemma~\ref{lem:type3} states that for every
   $p \leq k$ the number of $(j,b) \in \cal J$ such that $p_j \leq p$ is at
   most $p-1$.  Applying it for $p=k/(c^2\ln k)$ and $p=k$, as well as
   using~(\ref{ExPri: few large}) and~(\ref{ExPri: many small}), we get 
\[
   \sum_{(j,b) \in \cal J} 1/f(p_j) \leq \frac{k}{c^2\ln k} +
       \frac{k}{\ln k} = \frac{k}{\ln k} \left( 1+ \frac{1}{c^2} \right)
   \enspace.
\]

Putting things together, each job $i_0$ completed by the algorithm
receives a type 1 charge of at most $w_{i_0}$. By
Lemma~\ref{lem:type2} for $\rho=\alpha$ it can receive at most
$w_{i_0} k/ c^2 \ln k$ type 2 charges in total.  And we have just shown
that type 3 charges are, for large $k$, at most $w_{i_0}(1+1/c^2)k/\ln k$
in total.  Together, this is $w_{i_0}(1+2/c^2) \cdot k/\ln k
= w_{i_0}\left(3+o(1)\right) \cdot k/\ln k $.

It remains to prove the claims~(\ref{ExPri: few large}) and
(\ref{ExPri: many small}).  First let us observe that for every constant
$c<1$ and large enough $x$,
\begin{equation} \label{1/e bound}
    \left(1-\frac{c}{x}\right)^x \geq \frac 1 e \enspace,
\end{equation}
as for $x$ tending to infinity the left hand side tends to $e^{-c} > e^{-1}$.

Clearly $f(1)=1$ and, by (\ref{1/e bound}),
\begin{align*}
  f(k)	&= k \left(1-\frac{c^2\ln k}{k}\right)^{k-1}
		= k \left( 1-\frac{c^2\ln k}{k}\right)
		^{\frac{k}{c \ln k} (k-1) \frac{c \ln k}{k}} \\
  	&\geq k \left(\frac{1}{e}\right)^{(k-1)\frac{c \ln k}{k}}
		=k \cdot k^{c(1-k)/k} = k ^{(1-\epsilon+k\epsilon)/k}
		\geq \ln k \enspace,
\end{align*}
if $k$ is sufficiently large.

Now we observe that the sequence $(f(x))_{x=1}^k$ is non-decreasing
for $x\leq k/(c^2\ln k)$ and decreasing for $x > k/(c^2\ln k)$.  For
this we analyze the ratio $f(x)/f(x-1) = \alpha x/(x-1)$, and see
that it is at least $1$ if and only if $x \geq k/(c^2\ln k)$.
Inequalities (\ref{ExPri: few large}) and (\ref{ExPri:
many small}) follow.  This completes the proof.
\end{myproof}

\section{Identical Processing Times, upper bound} \label{sec: uniform-length}

In this section we consider instances where each job has the same
processing time $k\geq 2$ and arbitrary weight. 

\textsc{The Conservative Algorithm}: At every step execute the pending job
which maximises the priority $\pi(j,q_j) = 2^{-q_{j}/k} \cdot w_{j}$.

\begin{theorem}
The \CONSV{} is 5-competitive. 
\end{theorem}
\begin{myproof}
  The proof is based on a charging scheme, different from
  the general charging scheme of section~\ref{sec:intro}.

  Fix some instance.  Consider the jobs scheduled by the algorithm and
  jobs scheduled by the adversary.  Without loss of generality we
  assume that the adversary completes every job that he starts, and
  that he follows the \EDF{} policy.

  Every job $j$ scheduled by the adversary that is also completed by
  the algorithm, is charged to itself.  From now on we ignore
  those jobs, and focus on remaining ones.

  All jobs scheduled by the adversary will be charged to some jobs
  completed by the algorithm, in such a way that job $i$ completed by the
  algorithm receives a charge of at most $4w_i$ in total.

  For convenience we renumber the jobs completed by the algorithm from
  $1$ to $n$, such that the completion times are ordered
  $C_1<\ldots<C_n$.  Also we denote $C_0=0$.  For every $i=1,\ldots,n$
  we divide $[C_{i-1},C_{i})$ further into intervals: Let $a=\lceil
    (C_{i}-C_{i-1})/k \rceil$.  The first interval is
    $[C_{i-1},C_{i}-(a-1)k)$.  The remaining intervals are
      $[C_{i}-(b+1)k,C_{i}-bk)$ for every $b=a-2,\ldots,0$.  We label
        every interval $I$ with a pair $(b,i)$ such that $I=[s,C_{i}-bk)$
          for
          $s=\max\{C_{i-1},C_{i}-(b+1)k\}$.

           \begin{figure}
             \centerline{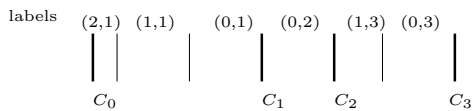}
             \caption{The intervals as used by the charging procedure.}
             \label{fig:blocks}
           \end{figure}
  
  The charging will be done by the following procedure, which
  maintains for every interval $[s,t)$ a set of jobs $P$ that are
    started before $t$ by the adversary and that are not yet charged
    to some job of the algorithm.

\begin{quote}
  Initially $P=\emptyset$.
  \\
  \textbf{For all} intervals $[s,t)$ as defined above  
                   in left to right order, do
  \begin{itemize}
    \item Let $(b,i)$ be the label of the interval.
  \item
     Add to $P$ all jobs $j$ started by the adversary in $[s,t)$.
  \item
     If $P$ is not empty, then remove from $P$ the job $j$ with
     the smallest deadline and charge it to $i$.  Mark $[s,t)$ with $j$.
  \item
     If $P$ is empty, then $[s,t)$ is not marked.
  \item
     Denote by $P_t$ the current content of $P$.
  \end{itemize}
\end{quote}

\begin{lemma}                                    \label{lem:pending}
   For every interval $[s,t)$, all jobs $j\in P_t$ are still pending
     for the algorithm at time $t$.
\end{lemma}
\begin{myproof}
Assume that $P_t$ is not empty, and let $j$ be the job in $P_t$ with
the smallest deadline.

First we claim that there is a time $s_0$, such that every
interval contained in $[s_0,t)$ is marked with some job $j'$
  satisfying $s_0 \leq r_{j'}$ and $d_{j'} \leq d_j$.

The existence of $s_0$ is shown by a kind of pointer chasing: Let
$[s',t')$ be the interval where the adversary started $j$.  So $j$
  entered $P$ by the charging procedure at this interval.  Job $j$ was
  in $P$ during all the iterations until $[s,t)$, so every interval
    between $t'$ and $t$ is marked with some job of deadline at most
    $d_j$.  Let $\cal M$ be the set of these jobs.  If for every
    $j'\in \cal M$ we have $s'\leq r_{j'}$, we choose $s_0=s'$ and we
    are done.  Otherwise let $j'\in\cal M$ be the job with smallest
    release time. So $r_{j'} < s'$.  Let $[s'',t'')$ be the interval
      where the adversary started $j'$.  By the same argment as above,
      during the iteration over the intervals between $s''$ and $s'$,
      job $j'$ was in $P$. Therefore every such interval was marked
      with some job with deadline at most $ d_{j'}\leq d_j$.  Now we
      repeat for $s''$ the argument we had for $s'$.  Eventually we
      obtain a valid $s_0$, since $P$ was initially empty.

Now let $\cal M$ be the set of jobs charged during all intervals in
$[s_0,t)$.  In an \EDF{} schedule of the adversary, job $j$ would
  complete not before $s_0 + (|{\cal M}|+1)k$.  But any interval has
  size at most $k$, so $t-s_0 \leq |{\cal M}| k$.  We conclude that
  $d_j \geq t+k$, which shows that $j$ is still pending for the
  algorithm at time $t$.
\end{myproof}

\begin{lemma}
    Let $[s,t)$ be an interval with label $(b,i)$ and $j$ a job
      pending for the algorithm at some time $t_0\in [s,t)$. Then $w_{j}
        \leq 2^{1-b} w_i$.
\end{lemma}
\begin{myproof}
Let $u=C_i$ and let $x_{t_0},x_{t_0+1},\ldots,x_{u-1}$ be the
respective priorities of the job units scheduled in $[t_0,u)$.
  Clearly the algorithm is $2^{-1/k}$-monotone, i.e.\ $x_{t'}\leq
  2^{-1/k} x_{t'+1}$ for every $t'\in[t_0,u)$.

      We have $x_{u-1} = 2^{-1/k} w_i$, since $i$ completes at $u$ and
      the remaining processing time of $i$ at time $(u-1)$ is 1.  Now
      the priority of $j$ at time $t_0$ is at most $2^{-1} w_j$,
      therefore
\[
    2^{-1}w_j \leq x_{t_0} \leq 2^{-(u-1-t_0)/k} x_{u-1} \leq 2^{-(u-t_0)/k} w_i = 2^{-b} w_i.
\]
\end{myproof}

This lemma permits to bound the total charge of a job $i$ completed by
the algorithm.  Let $a=\lceil (C_{i}-C_{i-1})/k \rceil$.  Then $i$
gets at most one charge of weight at most $2^{1-b}w_{i}$ for every
$b=a-1,\ldots,0$. Summing the bounds shows that job $i$ receives at
most $4$ times its own weight, plus one possible self-charge.

At time $t=C_n$ the algorithm is idle, so by Lemma~\ref{lem:pending},
$P_t=\emptyset$.  Therefore all jobs scheduled by the adversary have
been charged to some job of the algorithm, and this completes the
proof.
\end{myproof}

\section{Identical processing time, lower bound}   \label{sec:2.598}

%
%
\begin{theorem}
Any deterministic online algorithm for the equal processing time model with
$k \geq 2$ has competitive ratio at least $\frac{3}{2} \cdot \sqrt{3} \approx 2.598$.
\end{theorem}
\begin{myproof}
We describe the adversary's strategy for $k=2$ only, as it can be
easily adapted to larger values of $k$.  Every job $j$ will have
processing time $2$ and will be tight, i.e. $d_j=r_j+p_j=r_j+2$.
W.l.o.g. the adversary completes the heaviest feasible subset of jobs, which can
be specified once the sequence is finished. For the time being we need only
describe what jobs are released in each step. We also assume that when there are
pending jobs with positive weights, {\ALG} will process one of them, and that it
will never process a job with non-positive weight.

Initially ($t=0$) the adversary releases a job with weight $x_0=1$.
In every step $t>0$ the adversary releases a job with weight $x_t$ that we
specify later, unless the algorithm has already completed one job (this
has to be the one with weight $x_{t-2}$). In that case the adversary releases
no job at time $t$ and the sequence is finished. The adversary, in that case,
completes every other job starting from the last one, for a total gain of
$$ X_{t-1} = x_{t-1} + x_{t-3} + \ldots + x_{b+2} + x_{b} \enspace,$$
where $b=t-1 \bmod 2$, while {\ALG}'s gain is only $x_{t-2}$.

Now we describe the sequence $x_i$ that forces ratio at least
$R = 1.5\sqrt{3} - \epsilon$ for arbitrarily small epsilon. As we later prove,
there is a non-positive element $x_{i_0}$ in the sequence, so by previous
assumptions the algorithm completes some job released before the step $i_0$.

If {\ALG} completes a job released in step $t$, the ratio is
$$ R_t = \frac {X_{t+1}} {x_t} = \frac {X_{t+1}} {X_t-X_{t-2}} \enspace,$$
assuming $X_{-2}=X_{-1}=0$. As we want to force ratio $R$, we let $R_t=R$, i.e.
$$ X_{t+1} = R \left( X_t-X_{t-2} \right) $$
for each $t>0$. Note that this defines the sequence $x_i$, as
$x_i = X_i - X_{i-2}$.


To prove existence of $i_0$, we introduce two sequences:
$q_i = R \cdot X_{i-1} / X_{i+1}$ and $s_i = R-q_i = R(1-X_{i-1}/X_{i+1})$.
We shall derive a recursive formula
defining $q_i$ and $s_i$, and then prove that $s_i$ is a strictly decreasing
sequence. Next we prove that $s_i \leq 0$ for some $i$. That will conclude
the proof, as (assuming both $X_{i-1}$ and $X_{i+1}$ are positive)
$$ s_i \leq 0 \iff q_i \geq R \iff \frac {X_{i-1}} {X_{i+1}} \geq 1 \iff
	X_{i-1} \geq X_{i+1} \iff x_{i+1} \leq 0 \enspace. $$
Of course, if $X_{i-1} > 0$ and $X_{i+1} \leq 0$, then $x_{i+1}<0$ as well.

To prove existence of appropriate $i$ first observe that
\[
   X_i	= R \left( X_{i-1} - X_{i-3} \right)
	= X_{i-1} \left(R- R\frac {X_{i-3}} {X_{i-1}} \right)
	= X_{i-1} \left( R - q_{i-2} \right) \enspace,
\]
which implies
\begin{equation} \label{unilen-lb-q-rec}
   q_i = R \cdot \frac {X_{i-1}} {X_{i+1}} = \frac R {(R-q_{i-1})(R-q_{i-2})}
	\enspace.
\end{equation}
Rewriting~(\ref{unilen-lb-q-rec}) in terms of $s_i$ we get
\begin{equation} \label{unilen-lb-s-rec}
   s_i = R \left( 1- \frac 1 {s_{i-1}s_{i-2}} \right) \enspace,
\end{equation}
and one can calculate that $s_0=R$, $s_1=R-1/R$ and
$s_2=R(R^2-2)/(R^2-1)$, in particular $s_0>s_1>s_2>0$.

We prove by induction that $s_i$ is a decreasing sequence. Observe that
$$ s_{i+1}-s_i = R \left( \frac {1} {s_{i-1}s_{i-2}}
	- \frac{1} {s_i s_{i-1}} \right) = R \cdot \frac
	{s_i-s_{i-2}}{s_i s_{i-1} s_{i-2}} < 0 \enspace,$$
since by induction hypothesis $s_{i-2}>s_{i-1}>s_{i}$. Hence there is
$i$ such that $s_i \leq 0$, unless the sequence $s_i$ is bounded and
converges to $g = \inf s_i$, s.t. $g \geq 0$.  Suppose that is the case.
Then $s_i$ converges to $g$ and~(\ref{unilen-lb-s-rec}) holds for
$s_i=s_{i-1}=s_{i-2}=g$. Thus
$$ g = R \left( 1- \frac 1 {g^2} \right) \enspace, $$
or, equivalently,
\begin{equation}
   P(g) = g^3-Rg^2+R=0 \label{cubic-eq}
\end{equation}
Since $R = 1.5\sqrt{3} - \epsilon$, the discriminant of $P$, which is
$4R^2(R^2-27/4)$, is negative, i.e. $P$ has a single real root. As $P(-1)=-1$
and $P(0)=R>0$, the sole real root of $P$ lies in $(-1,0)$. In particular, it is
negative, which proves $s_i$ is not lower-bounded by any non-negative constant.
\end{myproof}

\paragraph{Discussion} The same construction was used
before~\cite{ChanLamTing:New-Results-on-On-Demand-Broadcasting}, and it was
claimed to yield $2.58$ lower bound on the competitive ratio. However, the proof
therein concludes with a statement that it can be verified that the sequence
$\{x_i\}$ contains a non-positive element $x_{i_0}$ if $R<2.58$.
In particular, the root $3\sqrt{3}/2$ appears neither in the theorem statement,
nor the proof. This, together with the fact that $i_0>60$ for $R=2.58-\epsilon$,
suggests that the claim of existence of $i_0$ was based on empirical results.

\section{Conclusion}

It remains open to determine the best competitive ratio a
deterministic algorithm can achieve for the equal processing time
model.  Even for $k=1$ the question is not completely answered.

How much the competitive ratio can be improved by use of randomization
remains unknown. The only paper~\cite{Canetti.Irani:randomized} we are
aware of studies the case of oblivious adversary and tight weighted jobs only.
It provides a lower bound of $\Omega(\sqrt{\log k/\log \log k})$
and an upper bound of $O(\log k)$ on the competitive ratio in that setting.
Can a similar ratio be achieved when jobs are not tight?

\vspace{0.5cm}
We would like to thank Artur Je\.z for his valuable comments.


\bibliographystyle{plain} 
\bibliography{onlinePreemptive}

\newpage
\appendix



\section{Bounded processing time, unit weights}

In this section we consider instances in which every job has 
processing time at most $k$ and unit weight, i.e $w_{i} = 1$ for all jobs $i$.

   \textsc{The} \SRPT{} Algorithm is a greedy online algorithm that schedules
   at every step the pending job with the smallest remaining processing time.

   It was analyzed in~%
   \cite{KalyanasundaramPruhs:Maximizing-job-completions-online},
   but we provide a concise proof, for completeness, using our general charging 
   scheme.
 \begin{proposition}[\cite{KalyanasundaramPruhs:Maximizing-job-completions-online}]
   \SRPT{} is $2H_k$-competitive, where $H_k$ denotes the $k$-th
   harmonic number, $1+1/2+1/3+\ldots+1/k$.
 \end{proposition}
 \begin{myproof}
   We use our general charging scheme.  The algorithm is 
   $\frac{k-1}{k}$-monotone and valid w.r.t. $\pi(i,a)=1/a$.
   Observe that whenever the algorithm schedules some job $i$
   at time $t$, then some job will
   complete in $[t+1,t+k+1)$, either $i$ itself or some job with smaller
   processing time. In particular if $t_0$ is the completion time of
   some job $i_0$ by the algorithm, and $s$ is the smallest time such that
   $[s,t_0)$ contains no idle time nor completion, then $t_0-s<k$
     and the unit scheduled at time $t_0-i$ for $1\leq i \leq t_0-s+1$
     has capacity at most $1/i$.  As a result the total type 2 charge
     to $i_0$ is at most $H_k$.

     Lemma~\ref{lem:type3} states that there are at most $p-1$ type 3
     charges to $i_0$ from jobs units $j$ with $p_j\leq p$.  The worst
     case is when there is exactly one job unit $j$ with $p_j=p$
     charging $1/p$ to $i_0$ for every $p=2,3,\ldots,k$.  Therefore the
     total type 3 charge to $i_0$ is at most $H_k-1$.

     Total type 1 charge is at most $w_{i_0}=1$,  so this concludes
     the proof.
 \end{myproof}

Now we prove an almost matching lower bound. Our construction is very similar
to one known before~\cite{BaruahHaritsaSharma:On-line-scheduling-to-maximize},
but the constant we obtain is slightly better.

\begin{theorem}
  Any deterministic online algorithm has ratio $\Omega(\log k/\log\log
  k)$.
\end{theorem}
\begin{myproof}
 Fix some deterministic algorithm.  We will define an instance denoted
 $I(\ell,0,0)$ from which the algorithm can complete at most a single
 job, and the adversary can complete $\ell$ jobs.  Moreover all jobs
 have processing time at most $(\ell+1)!$.  So if we choose
 $\ell = \lfloor \ln k/ \ln\ln k \rfloor -1$, the processing time
 is at most
\begin{align*}
(\ell+1)! &= \left\lfloor\frac {\ln k} {\ln \ln k}\right\rfloor!
		\leq \left(\frac {\ln k} {\ln \ln k}\right)^{\frac{\ln k}{\ln\ln k}} \\
	&= \exp\left((\ln\ln k - \ln\ln\ln k) \cdot \frac{\ln k}{\ln\ln k}\right) \leq \exp(\ln k) = k.
\end{align*}

Let $\ell \geq 1, s,e\geq 0$ be integers.  Let $f$ be a function
defined as $f(1,e)=e+1$ and for $\ell>1$,
\begin{equation} \label{lower-bound-span-formula}
   f(\ell,e) =  \max\{e,f(\ell-1,0)\} + f(\ell-1,0) 
                + f(\ell-1, \max\{e,f(\ell-1,0)\}) \enspace .
\end{equation}
We construct an instance $I(\ell,s,e)$ with the following properties.
\begin{itemize}
  \item The adversary can schedule $\ell$ jobs from this instance.
  \item The algorithm can schedule at most one job from this instance,
    and if it does, then it spends more than $e$ units on 
    jobs from this instance, including uncompleted ones.
  \item All jobs $i$ from the instance satisfy $s\leq r_i$ and
    $d_i\leq s+f(\ell,e)$, and therefore also $p_i\leq f(\ell,e)$.
\end{itemize}

The basis case is easy, for $I(1,s,e)$ at time $s$ we release a tight
job of length $e+1$.  It satisfies the required properties.

Now we show how to construct $I(\ell+1, s,e)$.  Let $b=f(\ell,0)$,
$a=\max\{e,b\}$ and $c=f(\ell,a)$.  At time $s$ we release a job $A$
of length $a+c$ and deadline $s+a+b+c$, as well as a job $B$ of length
$a+b$ and tight deadline.  At time $s+a$, if the algorithm scheduled
only $B$ in $[s,s+a)$, then we release instance $I(\ell,s+a,0)$.
  Otherwise at time $s+a+b$ we release $I(\ell,s+a+b, a)$, see
  Figure~\ref{fig:OmegaLogK}.

\begin{figure}[ht]
  \centerline{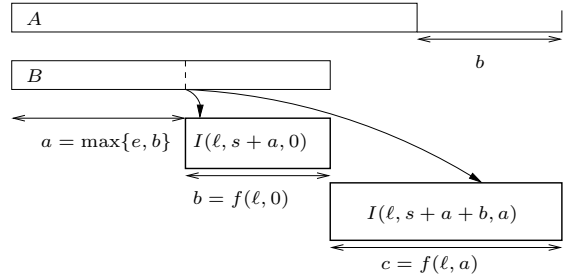}
  \caption{The construction of $I(\ell+1,s,e)$}
  \label{fig:OmegaLogK}
\end{figure}

Let us verify that the construction satisfies the required properties,
by induction on $\ell$.  We already settled the basis case $\ell=1$,
so assume the claim holds for instances $I(\ell,s',e')$ for all
$s',e'\geq 0$, and we will show it holds for $I(\ell+1,s,e)$ as well.
By construction and induction each job $i$ from instance
$I(\ell+1,s,e)$ is not released before $s$ and its deadline does not exceed
$s+a+b+c = s + f(\ell+1,e)$, so the third property is satisfied.
\begin{description}
  \item[In case the algorithm scheduled only $B$ in $[s,s+a)$:] At
    this point, if the algorithm completes $A$ or $B$, then in the
    interval $[s+a,s+a+b)$ there is not a single idle step left for
      another job.  Therefore by induction hypothesis the algorithm
      can only schedule a single job.  The algorithm already spent $a$
      units on $B$, so if it does complete a job, then it spends
      more than $a \geq e$ units on jobs from this
      instance. By induction hypothesis, the adversary can schedule
      $\ell$ jobs from the subinstance in the interval
      $[s+a,s+a+b)$, and schedule $A$ in the remaining time units
        $[s,s+a) \cup [s+a+b,s+a+c)$.
      \item[Otherwise:] The algorithm cannot complete $B$, since the
  job is tight.  If the algorithm completes some job from
  $I(\ell,s+a+b,a)$, then by induction hypothesis, it spends strictly
  more than $a\geq e$ units on jobs from the sub-instance.  This does
  not leave enough space to complete job $A$ in addition.  And if the
  algorithm completes job $A$, it spends $a+c >e$ units on it.
  The adversary can complete $B$ plus $\ell$ jobs from the
  sub-instance.
\end{description}

To complete the proof of the theorem, 
it remains to show that all jobs from $I(\ell,0,0)$ have processing
time at most $(\ell+1)!$.  To this end, we prove by induction that
\begin{equation} \label{lower-bound-span-formula-solved}
   f(\ell,e) =	\ell \max\left\{ \ell!, (\ell-1)! + e \right\},
\end{equation}
which implies that all jobs from $I(\ell,0,0)$ have processing time at
most $\ell \cdot \ell ! < (\ell+1)!$.
Note that~(\ref{lower-bound-span-formula-solved}) trivially holds for
$\ell = 1$. Now assume it holds for $\ell - 1$, and in particular
$f(\ell-1,0)=(\ell-1)(\ell-1)!$.  Then
\begin{align}
f(\ell,e) 	&= \max\{e,f(\ell-1,0)\} + f(\ell-1,0) + f(\ell-1, \max\{e,f(\ell-1,0)\}) \notag\\
		&= \max\{ e, (\ell-1) \cdot (\ell-1)! \} + (\ell-1) \cdot (\ell-1)! + \notag\\
		& 	\qquad + ( \ell-1 ) \max\{ (\ell-1)!,
			(\ell-2)! + \max\{ e, (\ell-1) \cdot (\ell-1)! \} \} \notag\\
		&= \max\{ e, (\ell-1) \cdot (\ell-1)! \} + (\ell-1) \cdot (\ell-1)! + \notag\\
		& 	\qquad + ( \ell-1 )\cdot (
			(\ell-2)! + \max\{ e, (\ell-1) \cdot (\ell-1)! \}) \label{eq:l2}\\
		&= \ell \cdot \max\{ e,(\ell-1) \cdot (\ell-1)! \} +\ell! \notag\\
                &= \ell \cdot \max\{e+(\ell-1)!, \ell!\} \notag
\end{align}
The equality (\ref{eq:l2}) follows from $(\ell-1)! <
(\ell-2)! + \max\{ e, (\ell-1) \cdot (\ell-1)! \}$.  
\end{myproof}

\section{Lower Bound for bounded processing times}

%
%
Ting~\cite{Ting:broadcasts} showed that competitive ratio of any deterministic
algorithm in this setting is at least $k/(2\ln k)-1$, while we improve it to
$k/\ln k -o(1)$.

\begin{lemma}		\label{lem:log}
For any deterministic algorithm  its competitive ratio is at least
$k/\ln k - o(1)$. In particular, it is at least $k/\ln k - 0.06$ for
$k \geq 16$.
\end{lemma}
\begin{myproof}
  For convenience denote $R=k/\ln k$, $r=\lceil R \rceil -1$, and assume
  $k\geq 16$.  Fix any deterministic algorithm and consider the following
  instance, depicted in Figure~\ref{fig:KOverLogK}.  At
  time 0, the adversary releases a big job $B$ with weight $w_{B} =
  R$, processing time $k$ and deadline $k$, as well as a
  small job $A_1$ with weight, processing time and deadline all
  $1$. Moreover, at each moment $0 \leq t \leq k-1$, if the algorithm
  scheduled only job $B$ in $[0,t)$, then the adversary releases a
  tight job $A_{t+1}$ of unit processing time at time $t$, and does
  not release any new job otherwise.  The jobs $A_{t}$ have weights:
\[
w(A_{t}) := \begin{cases}
			1 & \text{ if } t < R \enspace, \\
			e^{t/R-1} & \text{ if }  t\geq R \enspace.
		 \end{cases} 
\]    
Note, job $A_t$ is released at time $t-1$.

\begin{figure}[ht]
  \centerline{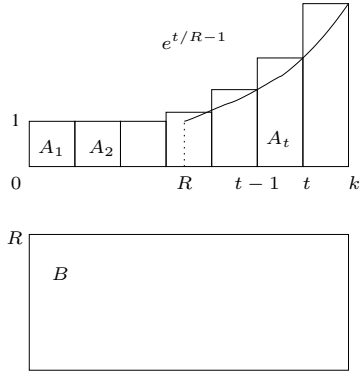}
  \caption{The construction of the lower bound}
  \label{fig:KOverLogK}
\end{figure}

If the algorithm schedules a job $A_{t_0}$ with $t_0<R$, then the adversary
schedules job $B$ and the ratio is $R$.

If the algorithm schedules a job $A_{t_0}$ with $t_0 \geq R$, then the
adversary schedules all jobs $A_t$ for $t=1,\ldots, t_0$.  The
adversary's gain is 
\begin{align}
   \lceil R \rceil-1 +\sum_{t=\lceil R\rceil}^{t_0} e^{t/R-1}
	&= r + \sum_{t=r+1}^{t_0} e^{t/R-1}
	\geq r + \int_{r}^{t_0} e^{t/R-1} \mathrm{d} t  \nonumber \\
&= r + \left[ R e^{t/R-1} \right]_{r}^{t_0}
	= r - Re^{r/R -1} + R e^{t_0/R-1} \nonumber \\
&= f(R,r) + R e^{t_0/R-1} = f(R,r) + R w(A_{t_0}) \enspace,
	\label{prefix-sum}
\end{align}
where the inequality follows from monotonicity of the function $e^{t/R-1}$, and
$$ f(R,r) := r - Re^{r/R -1} \enspace.$$
So the adversary gain is at least $k/\ln k$ times the algorithm's
gain plus $f(R,r)$. 

If the algorithm schedules job $B$, gaining $k/\ln k$, the adversary
schedules all $k$ jobs $A_t$ from $t=0$ to $k-1$.  In that case,
by~(\ref{prefix-sum}) its gain is at least
\[
    f(R,r) + R e^{k/R-1} =  f(R,r) + R e^{\ln k - 1} = f(R,r) + R \cdot k/e,
\]
and we need it to be more than $f(R,r) + R w(B) = f(R,r) + R^2$.  This is true
if $e \leq \ln k$ which holds for $k \geq e^e$, in particular when $k\geq 16$.  

Now we analyze the function $f(R,r)$. Recall that $R = k/\ln k$ and
$r = \lceil R \rceil -1$, so in particular $R-r \in (0,1]$.
As $e^x \geq 1+x$ and both sides converge to $1$ as $x$ tends to $0$, we have
\[
  f(R,r) = r - Re^{r/R -1} \leq r - R \cdot \frac{r}{R} = 0 \enspace,
\]
and $f(R,r)$ tends to $0$ as $k$ grows.

In particular, it is straightforward to check that $f\left(R(k),r(k)\right)
\geq -0.06$ for $k=16,17,\ldots,21$, and that $f\left(R(k),r(k)\right) \geq
f(7,1) > -0.06$ for larger $k$.  As the algorithm's gain is (w.l.o.g.)
at least $1$, $f(R,r)$ divided by that gain is at least $f(R,r)$, which
concludes the proof.
\end{myproof}


\end{document}